\documentclass[aps,prb,twocolumn]{revtex4} 
\usepackage{graphicx}

\begin{document}

\newlength\FigWidth 
\FigWidth 3.25 true in

\title{Spin Waves in Disordered III-V Diluted Magnetic Semiconductors}

\author{Mona Berciu$^1$ and R. N. Bhatt$^{1,2}$}

\affiliation{ $^1$Princeton Materials Institute, Princeton University,
Princeton, NJ 08544, USA \\ $^2$Department of Electrical Engineering,
Princeton University, Princeton, New Jersey 08544-2163}

\date{\today}

\begin{abstract}
We propose a new scheme for numerically computing collective-mode
spectra for large-size systems, using a reformulation of the Random
Phase Approximation.  In this study, we apply this method to
investigate the spectrum and nature of the spin-waves of a (III,Mn)V
Diluted Magnetic Semiconductor. We use an impurity band picture to
describe the interaction of the charge carriers with the local Mn
spins. The spin-wave spectrum is shown to depend sensitively on the
positional disorder of the Mn atoms inside the host
semiconductor. Both localized and extended spin-wave modes are
found. Unusual spin and charge transport is implied.
\end{abstract}


\maketitle

\section{Introduction}

Diluted Magnetic Semiconductors (DMS) based on III-V semiconductors
doped with Mn have attracted a lot of interest recently, after
critical temperatures for the onset of ferromagnetism of the order of
110 K have been found in Ga$_{1-x}$Mn$_x$As, for
$x=0.053$.\cite{Ohno,Haya,Esch} More recently, critical temperatures
larger than room temperatures have been reported in Mn-doped GaN,
enhancing the hope for extensive technological applications of these
materials.\cite{Hebard,Sono}

While there is general agreement that the magnetism is due to
charge-carrier mediated, effectively ferromagnetic interactions
between the Mn spins, there are various theoretical models attempting
to understand their detailed behavior. The DMS are alloy systems, with
inherent positional disorder of the Mn atoms; further, spin-orbit
coupling may play a significant role for hole doping. A theoretical
treatment which takes into account all these factors, and their
effects on the magnetic and transport properties of DMS, is not yet
available. Instead, theoretical models have tended to concentrate on
different aspects of the problem.

One class of models, where much work has been done, neglects both the
disorder and spin-orbit coupling effects.  At large carrier densities,
where Coulomb potentials of the impurity (Mn) atoms are effectively
screened out, and other disorder effects (such as those coming from As
antisite defects which are believed to be the cause of the large
compensation observed experimentally) can be neglected, the former at
least can be justified. In such a case, holes occupy a Fermi sea at
the top of the valence band.\cite{RKKY} Studies of the spin-wave
spectra\cite{MacD}, Monte-Carlo studies\cite{MacDonald} as well as
dynamical mean-field theory studies\cite{Millis} have been
performed. Overall, their results are rather similar to the physics
one would expect to hold in a conventional ferromagnet. More recently,
it has been proposed that spin-orbit coupling of the valence band, in
the high carrier density (strongly metallic) regime leads to an RKKY
coupling between Mn moments that is {\em anisotropic} in spin space,
and where the positional disorder effectively leads to random
anisotropy.\cite{Janko} This would lead to frustration effects on the
magnetic ordering, and therefore unusual ferromagnetism.  A number of
very useful ab-initio studies have also been published.\cite{Sanvito}

In parallel, we have studied effects of positional disorder on the
magnetic ordering (in the absence of spin-orbit coupling) of an
impurity band model,\cite{MB1,MB2,Malcolm,malcolm2} developed from
previous work done in the context of insulating II-VI DMS
compounds.\cite{Xin,MB4} Such an approach should be of relevance at
low doping concentrations x, at and below the metal-insulator
transition, and possibly even above.  Evidence for the existence of
impurity-like states is provided by ab-initio studies,\cite{Sanvito}
which found that occupied hole states near the Fermi level have
wave-functions mostly concentrated on and near the Mn impurity.  More
recently, Angle Resolved Photoemission Spectroscopy has revealed the
existence of the impurity band in Ga$_{1-x}$Mn$_x$As with $x=0.035$
(very close to the metal-insulator transition).\cite{Okabayashi} A
scanning tunneling microscope study \cite{STM} demonstrates the
existence of an impurity band in (Ga,Mn)As samples with
x=0.005-0.06. Optical spectroscopy\cite{Basov} also identifies the
impurity band for x=0.0001 and x=0.05 samples.

Electron densities in localized states, as well as states close to the
mobility edge, can be far from homogeneous, unlike Bloch waves.  At
low Mn concentrations and high degree of compensation (which comes
from charged centers) seen in DMS, short length-scale density
fluctuations could be significant.  This can lead to inhomogeneities
in the local charge densities at different Mn sites, which in turn
leads to (microscopically) inhomogeneous magnetizations of the
magnetic ions in the ordered phase. Such inhomogeneity would be
expected to alter the nature of the collective magnetic excitations
(spin waves) of the magnetically ordered system, which in turn would
affect charge transport through the magnetic excitation processes
which give rise to spin-flip scattering.

In this paper, we develop a numerical scheme to calculate the spin
wave spectrum for a finite but large size model of coupled fermions
and spins, in the presence of quenched disorder. We show that the
method accurately reproduces the results for a lattice model of small
size obtained from a standard treatment of spin waves via the Random
Phase Approximation.\cite{Negele,Rowe,Ripka} We then apply the method
to the model of Ref. \onlinecite{MB1}, and study in detail its
collective magnetic excitations.

The plan of the paper is as follows. The model Hamiltonian for DMS is
described in section II. Section III is devoted to developing the
numerical scheme. The accuracy of the scheme is tested by comparing it
with the standard RPA method in section IV. In particular, we
demonstrate that we can use our scheme to study large size systems,
beyond the capability of the standard RPA approach.  Section V
presents the results obtained by applying the scheme to the model of
Section II, including the density of states and nature of the
eigenvectors of the magnetic excitations of the coupled fermion-spin
system. We summarize our conclusions in section VI.

\section{The model}
\label{sec2}

In the following we develop the formalism for an impurity band
 Hamiltonian that we believe to be appropriate for describing (at
 least qualitatively) the diluted magnetic semiconductor
 Ga$_{1-x}$Mn$_{x}$As for low doping $x$ (in the insulating phase, or
 not too far from the metal-insulator transition).  However,
 generalizations of this method for other types of Hamiltonians,
 crystal structures, parameters etc.  can be done in a straightforward
 manner.
  
The III-V host semiconductor is assumed to have the zinc-blende
structure appropriate for GaAs. Experimental
measurements\cite{Ohnorev} suggest that valence-II Mn substitutes for
valence-III Ga. As a result, the $N_d$ Mn dopants are randomly
distributed at positions $\vec{R}_i$, $i=1,...,N_d$ on the $N\times N
\times N$ FCC Ga sublattice, corresponding to $x = N_d/4N^3$. All
throughout this paper, we assume periodic boundary conditions.

Due to the valence mismatch, each Mn introduces a charge carrier
(hole) in the system. However, experimentally it is found that the
hole concentration is considerably suppressed through compensation
processes. As a result, the number of holes is only a fraction
$p=10-30\%$ of the number of Mn atoms $N_h = pN_d$. Each Mn atom also
has a ${5 \over 2}-$spin $\vec{S}_i$ from its half-filled $3d$
level. The magnetic properties of the doped semiconductor are related
to the exchange interaction between the Mn spins and the charge
carrier spins. This interaction is known to be antiferromagnetic
(AFM), and proportional to the probability of finding the charge
carrier at the corresponding Mn site.

Recently, we proposed a simple impurity-band model to describe the
behavior of the charge carriers for the low-doping
regime.\cite{MB1,MB2} The main justification is that near and below
the metal-insulator transition ($x\approx 0.03$) the density of charge
carriers is not large enough to effectively screen the attractive
Coulomb potential of the Mn dopants. As a result, bound, hydrogen-like
impurity states are created about each Mn site, at an energy $E_b = 1$
Ry above the top of the valence band. Due to interactions, these
impurity states lead to the appearance of an impurity band, and the
holes first occupy states in this band. Only if the concentration of
holes (or the temperature) is large enough, are states in the valence
band itself occupied by holes. Thus, it is reasonable to attempt a
description of the charge carriers behavior only in terms of this
impurity band.

As in previous work,\cite{MB1,MB2} in the following we will use the
electron formalism to treat the problem. This is equivalent to
performing a particle-hole transformation which leads to an inversion
$E \rightarrow -E$ of the charge carrier spectrum.  Thus, instead of
emptying the top of a valence-like impurity band (i.e., introducing
holes in the system) we instead occupy the bottom of a conduction-like
impurity band. We also make the simplifying assumption that the
isolated impurity wave-function for a charge carrier trapped near the
Mn at site $\vec{R}_i$ is the $1s$ hydrogen orbital
$\phi(\vec{r}-\vec{R}_i)=\phi_i(\vec{r}) \sim
\exp{(-|\vec{r}-\vec{R}_i|/a_B )}$, where $a_B=\hbar/\sqrt{2m^*E_b}=8$
\AA ~is the appropriate Bohr radius related to the effective mass of
the heavy hole ($m^*\approx 0.5m_e$ for GaAs) and the binding energy
($E_b = 112.4$ meV for Mn in GaAs).\cite{BG} Our approach neglects
both the complicated orbital form of the acceptor wave-function (Ref.
\onlinecite{BG}) and spin-orbit coupling. The former is not expected
to lead to any qualitative changes. It has recently been proposed that
spin-orbit coupling leads to frustration in the magnetic
ordering.\cite{Janko} These effects are left out in the present study,
which concentrates on the effect of disorder. A study combining both
effects is indicated for future work.

We consider the Hamiltonian
\begin{equation}
\label{1.1}
{\cal H} = \sum_{i,j,\sigma}^{}t_{ij} c^{\dagger}_{i\sigma}
c_{j\sigma} + \sum_{i,j}^{}J_{ij} \vec{S}_i \cdot \vec{s}_j.
\end{equation}
Here, $c^{\dagger}_{i\sigma}$ creates a charge carrier with spin
$\sigma$ in the impurity state centered at site $\vec{R}_i$. The first
term describes the hopping of charge carriers between impurity
states. We use the parameterization $t_{ij} = 2(1+r/a_{\rm
B})\exp{(-r/a_{\rm B})}$ Ry, where $r=|{\bf R}_i - {\bf R}_j|$, of
magnitude and form appropriate for hopping between two isolated 1$s$
impurities which are not too close to one another. \cite{Bhatt1} This
hopping term has been shown to lead to the appearance of an impurity
band which has a mobility edge, as well as a characteristic energy for
the occupied states in agreement with physical expectation.\cite{MB3}

The second term of the Hamiltonian (\ref{1.1}) describes the AFM
exchange between the Mn spin ${\vec S}_i$ and the charge carrier spin
${\vec s}_j={ 1\over 2} c^{\dagger}_{j\alpha} {\vec
\sigma}_{\alpha\beta} c_{j\beta}$ [${\vec \sigma}$ are the Pauli spin
matrices]. This AFM exchange is proportional to the probability of
finding the charge carrier trapped at ${\vec R}_j$ near the Mn spin at
${\vec R}_i$, and therefore $J_{ij} = J |\phi_j(\vec{R}_i)|^2 = J
\exp{\left(-2 |{\vec R}_i-{\vec R}_j|/a_B \right)}$. Based on
calculations of the isolated Mn impurity in GaAs, we estimate the
exchange coupling between a hole and the trapping Mn
($\vec{R}_i=\vec{R}_j$) to be $J=15$ meV.\cite{MB1,BG} As already
stated, the number of Mn atoms is $N_d$, and therefore there are a
total of $2N_d$ states in the impurity band.  The number of charge
carriers is fixed to $N_h=pN_d$, where we take $p=10\%$.

In Ref. \onlinecite{MB2} we studied the relevance of various other
terms, such as an on-site potential (associated with the Coulomb
potential of the compensation centers), Hubbard-like on-site repulsion
(describing interactions between charge-carriers), external magnetic
field etc. While they lead to various {\em quantitative} changes, we
believe that the {\em qualitative} picture we present in the following
sections is not changed by their absence.

\section{The collective mode spectrum}
\label{sec3}

The Random Phase Approximation (RPA) describes the collective
excitations of a system about its self-consistent $T=0$ mean-field
ground state. In order to clarify the notation used, we begin this
section with a very short review of the derivation of the relevant
equations for the mean-field ground state.

\subsection{The mean-field ground state}

We use the equation-of-motion approach of Ref. \onlinecite{Rowe}. At
the mean-field level, we are trying to find non-interacting fermionic
quasiparticles through a unitary transformation of the on-site charge
carrier operators:
\begin{equation}
\label{21}
a^{\dagger}_{n\sigma} = \sum_{i}^{} \psi_{n\sigma}(i)
c^{\dagger}_{i\sigma}
\end{equation}
such that
\begin{equation}
\label{22}
[{\cal H}, a^{\dagger}_{n\sigma}] \approx E_{n\sigma}
a^{\dagger}_{n\sigma}
\end{equation}
and
\begin{equation}
\label{23}
[{\cal H}, S_i^{+}] \approx - \hbar H_i S_i^{+}
\end{equation}
For a non-interacting system, Eqs. (\ref{22}) and (\ref{23}) would be
exact. For an interacting system at the mean-field level, the exact
Hamiltonian is approximated by the diagonalized, non-interacting
Hartree-Fock Hamiltonian
\begin{equation}
\label{24}
{\cal H} \rightarrow {\cal H}_{HF} = \sum_{n\sigma}^{} E_{n\sigma}
a^{\dagger}_{n\sigma} a_{n\sigma} - \sum_{i}^{} H_i S^z_i
\end{equation}
for which Eqs. (\ref{22}) and (\ref{23}) become exact.

It is straightforward to show that for Hamiltonian (\ref{1.1})
\begin{equation}
\label{25}
[{\cal H}, a^{\dagger}_{n\sigma}] = \sum_{ij}^{} t_{ij}
\psi_{n\sigma}(j) c^{\dagger}_{i\sigma} + \sum_{ij}^{} J_{ij}
\vec{S}_j {\vec{\sigma}_{\alpha\sigma} \over 2} \psi_{n\sigma}(i)
c^{\dagger}_{i\alpha}
\end{equation}
The right-hand-side of Eq. (\ref{25}) contains spin operators, so it
cannot be put in the form of Eq. (\ref{22}) unless we linearize it, by
replacing the spin operators with their mean-field ground state
expectation value $\vec{S}_j \rightarrow \langle \vec{S}_j\rangle =
\hat{e}_z S_{Mn}(j)$. While the choice of a collinear ground-state,
with all Mn spins aligned along the $z$-axis, is not the most general
possibility, we have actually shown in Ref. \onlinecite{MB2} that the
ground-state of this model, for the range of parameters we are
interested in, is indeed collinear. This is why we choose to apriori
make this assumption in this case.

After the linearization, we can now equate the right-hand-side of
Eq. (\ref{25}) with $E_{n\sigma} a^{\dagger}_{n\sigma}$ [see
Eq. (\ref{22})] to obtain the Hartree-Fock equation for the
one-particle orbitals:
\begin{equation}
\label{26}
E_{n\sigma} \psi_{n\sigma}(i) = \sum_{j}^{}t_{ij} \psi_{n\sigma}(j) +
\sum_{j}^{} J_{ij} S_{Mn}(j) { \sigma \over 2} \psi_{n\sigma}(i)
\end{equation}
The mean-field ground-state of the charge carriers is obtained by
occupying the $N_h$ lowest-energy orbitals,
\begin{equation}
\label{27}
| \Psi\rangle_{cc} = \prod_{(n\sigma)=1}^{N_h}
  a^{\dagger}_{n\sigma}|0\rangle
\end{equation}

The effective magnetic fields $H_i$ [see Eq. \ref{24})] are obtained
in a similar way. We evaluate the commutator of the exact Hamiltonian
with the spin raising operator $S_i^{+} = S_i^x + i S_i^y$
\begin{equation}
\label{28}
[{\cal H}, S_i^{+}] = -\hbar \sum_{j}^{}J_{ij} S_i^z s_i^{+} + \hbar
\sum_{j}^{} J_{ij} S_i^{+} s_i^{z}
\end{equation}
The right-hand-side now contains charge carrier operators (in the
charge carrier spin operator), beside the spin operators.  We again
use linearization $s_i^{+} \rightarrow \langle s_i^{+} \rangle =0$
[see Eq. (\ref{27})], $s_i^{z} \rightarrow \langle s_i^{z} \rangle =
s_h(i)$, and by comparison with Eq. (\ref{23}) we find
\begin{equation}
\label{29}
H_i = - \sum_{j}^{}J_{ij} s_h(j)
\end{equation}

From Eq. (\ref{24}) we see that if $H_i > 0$, then in the mean-field
ground state the spin $\vec{S}_i$ is in the ``up'' state $|+S \rangle$
and therefore $S_{Mn}(i) = +S$, while if $H_i < 0$ the spin
$\vec{S}_i$ is in the ``down'' state $|-S\rangle$ and $S_{Mn}(i)=-S$
($S=5/2$). As a result, the spin-part of the mean-field ground-state
can also be easily found if the expectation value of the charge
carrier spins $s_h(j) $ are known from Eqs. (\ref{26}),
(\ref{27}). Thus, we obtain the usual self-consistent Hartree-Fock
equations.

Diluted Magnetic Semiconductors exhibit ferromagnetism at low
temperatures. As a result, we assume that in the mean-field ground
state, all Mn spins are fully polarized $S_{Mn}(i) = +S$. Then, from
Eq. (\ref{26}) we find the lowest-lying $N_h$ states. If the charge
carriers are also fully polarized at $T=0$, all the occupied $N_h$
state have $\sigma = \downarrow$, and therefore the mean-field
ground-state is of the form:
\begin{equation}
\label{30}
| \Psi\rangle = \prod_{n=1}^{N_h} a^{\dagger}_{n\downarrow} | 0
  \rangle \otimes |S,S,....,S\rangle
\end{equation}

Of course, one must check that self-consistency is obeyed by verifying
that indeed the first $N_h$ lowest energy one-electron states are all
spin-down. We find that this always holds true for the parameters we
study in this paper. As discussed later, the spin-wave spectrum of the
collective excitations also confirms that this ground-state is indeed
stable.  However, for higher charge carrier concentrations (or
Hamiltonians with other types of interactions) it is likely that the
HF ground-state is only partially spin polarized. In that case, one
must do the full iterational search for the self-consistent HF ground
state. The computation for the spin-wave spectrum in the
partially-polarized case (for instance due to spin-orbit coupling),
can be derived in a similar way to the one we present in the following
for the fully polarized case.

\subsection{The Random Phase Approximation}

We will use the same equation-of-motion approach to derive the RPA
equations. We are interested in spin-waves, which are related to
spin-flip (spin-lowering) processes. Therefore, the RPA operators for
spin-wave collective modes must be constructed in terms of spin-flip
operators
\begin{equation}
\label{a.1}
\left\{
\begin{array}[c]{c}
b^{\dagger}_{hp} = a^{\dagger}_{h\uparrow} a_{p\downarrow} \\
B^{\dagger}_i = {1 \over \sqrt{2S}} S^-_i \\
\end{array}
\right.
\end{equation}
The index $h=1,N_d$ runs over the empty (``hole'') states of the HF
ground state, while the index $p=1,N_h$ runs over the occupied
(``particle'') states of the HF ground state. The index $i=1,N_d$ runs
over all the Mn spin positions. The HF ground-state $|\Psi\rangle$ is
fully polarized, and given by Eq. (\ref{30}).

The spin-flip operators have ``bosonic'' nature, as expected for
collective modes operators. Indeed, it is straightforward to verify
that $\langle \left[ b_{hp}, b^{\dagger}_{h'p'}\right]\rangle =
\delta_{pp'} \delta_{hh'}$, $\langle \left[ B_i,
B^{\dagger}_j\right]\rangle = \delta_{ij}$, with all the other
commutators identically zero. Here, the notation $\langle ... \rangle$
signifies an average over the HF ground state $ \langle \Psi | ... |
\Psi\rangle$.

The Hamiltonian can be rewritten in terms of the $a_{n\sigma}$
operators as [see Eqs. (\ref{1.1}), (\ref{21})]
\begin{equation}
\label{a.2}
{\cal H}=\sum_{n,m,\sigma}^{}t_{nm\sigma}a^{\dagger}_{n\sigma}
a_{m\sigma} + \sum_{n,m,i \atop \alpha\beta}^{} J_{n\alpha,m\beta}(i)
a^{\dagger}_{n\alpha} a_{m\beta} \vec{\sigma}_{\alpha\beta}\cdot
\vec{S}_i
\end{equation}
where
$$ t_{nm\sigma} = \sum_{i,j}^{}t_{ij}
\psi^*_{n\sigma}(i)\psi_{m\sigma}(j)
$$ and
\begin{equation}
\label{31}
J_{n\alpha, m\beta}(i) = { 1\over 2} \sum_{j}^{} J_{ij}
\psi_{n\alpha}^*(j) \psi_{m\beta}(j).
\end{equation}

One can easily derive the equations of motion for the spin-flip
operators. For instance
$$ [{\cal H}, B^{\dagger}_i]= { 2 \over \sqrt{2S}} \sum_{nm}^{}
J_{n\downarrow, m\uparrow}(i) a^{\dagger}_{n\downarrow} a_{m\uparrow}
S^z_i
$$
\begin{equation}
\label{a.3}
- { 1 \over \sqrt{2S}} \sum_{nm\sigma}^{} J_{n\sigma,m\sigma}(i)
\sigma a_{n\sigma}^{\dagger} a_{m\sigma} S^-_i
\end{equation}
Since RPA is an approximation, not an exact solution, we again must
linearize this commutator about the HF ground-state, and keep only the
meaningful terms. In other words, we replace
$$ a^{\dagger}_{n\downarrow} a_{m\uparrow} S^z_i \rightarrow \langle
a^{\dagger}_{n\downarrow} a_{m\uparrow} \rangle S^z_i +
a^{\dagger}_{n\downarrow} a_{m\uparrow} \langle S^z_i \rangle
\rightarrow S a^{\dagger}_{n\downarrow} a_{m\uparrow}
$$ Here, the first transformation is the linearization about the HF
ground state $|\Psi\rangle$. The second transformation results if one
uses the HF expectation values $\langle S^z_i \rangle = S$, $\langle
a^{\dagger}_{n\downarrow} a_{m\uparrow}\rangle=0$. The restriction to
relevant terms means that in any sum over $(n\downarrow)$, we restrict
$n$ to being an occupied orbital in the HF ground-state $1\le n \le
N_h$, because otherwise $b_{mn}^{\dagger} |\Psi\rangle = 0$, i.e. a
spin-flip process from the HF ground-state is impossible [see
Eq. (\ref{a.1})].  Performing a similar linearization for
$a_{n\sigma}^{\dagger} a_{m\sigma} S^-_i$ and collecting the various
terms, we find the linearized equation of motion
\begin{equation}
\label{a4}
[{\cal H}, B^{\dagger}_i]\approx H_i B_i^{\dagger} + \sqrt{2S}
\sum_{p=1}^{N_h} \sum_{h=1}^{N_d} J_{p\downarrow, h\uparrow}(i) b_{hp}
\end{equation}
After similar steps, the linearized equation of motion for the other
spin-flip operator $b_{hp}$ is found to be
\begin{equation}
\label{a5}
[{\cal H}, b_{hp}] \approx -\left( E_{h\uparrow} -
E_{p\downarrow}\right) b_{hp} - \sqrt{2S} \sum_{i}^{}
J^*_{p\downarrow, h\uparrow}(i) B_i^{\dagger}
\end{equation}

Eqs. (\ref{a4}) and (\ref{a5}) allow us to ``guess'' the RPA part of
the Hamiltonian, for which Eqs. (\ref{a4}) and (\ref{a5}) become
exact, to be given by
$$ {\cal H}_{RPA} = \sum_{h=1}^{N_d} \sum_{p=1}^{N_h}
\left(E_{h\uparrow} - E_{p\downarrow}\right) b^{\dagger}_{hp}b_{hp} +
\sum_{i}^{}H_i B^{\dagger}_i B_i
$$
\begin{equation}
\label{a.6}
+\sqrt{2S}\sum_{i}^{}\sum_{h=1}^{N_d} \sum_{p=1}^{N_h}\left(
J_{p\downarrow, h\uparrow}(i) B_i b_{hp} + h.c.\right)
\end{equation}
Thus, the RPA Hamiltonian is quartic in electron operators and
quadratic in spin operators, showing that it is the next order term
after the Hartree-Fock Hamiltonian (which is quadratic in electron
operators and linear in spin operators)
$$ {\cal H} = {\cal H}_{HF} + {\cal H}_{RPA} + ....
$$ Higher order approximation terms describe interactions between the
collective modes, which are neglected at the RPA level.

We want to diagonalize the RPA Hamiltonian to the canonical form
\begin{equation}
\label{a.7}
{\cal H}_{RPA} = \sum_{\alpha}^{}\hbar\Omega_{\alpha}
Q^{\dagger}_{\alpha} Q_{\alpha}
\end{equation}
where $Q^{\dagger}_{\alpha}$ is the creation operator for a spin-wave
with energy $\hbar\Omega_{\alpha}$, and $\alpha$ is an index (in
homogeneous systems, it would be a wave-vector). Since they create
collective mode excitations, these operators must have bosonic nature
$[Q_{\alpha}, Q_{\beta}]=0$. However, since they do not describe
exact, but only approximative solutions of the exact Hamiltonian, in
fact only the weaker condition $\langle[Q_{\alpha},
Q^{\dagger}_{\beta}]\rangle=\delta_{\alpha\beta}$ is obeyed. The most
general form for these operators [see Eq. (\ref{a.6})] is
\begin{equation}
\label{a.8}
Q^{\dagger}_{\alpha}= \sum_{h=1}^{N_d}\sum_{p=1}^{N_h}
X_{hp}^{(\alpha)} b_{hp} + \sum_{i}^{} Y_i^{(\alpha)} B^{\dagger}_i
\end{equation}
Using the equation of motion $[{\cal H}_{RPA}, Q^{\dagger}_{\alpha}]=
\hbar \Omega_{\alpha} Q^{\dagger}_{\alpha} $ and computing the
commutator using Eqs. (\ref{a.6}), (\ref{a.8}), we find the
diagonalization condition to be
\begin{eqnarray}
\label{a.9}%
&& \left[ \hbar\Omega_{\alpha} + E_{h\uparrow} -
E_{p\downarrow}\right] X^{(\alpha)}_{hp} = \sqrt{2S}
\sum_{i}^{}J_{p\downarrow,h\uparrow}(i) Y^{(\alpha)}_i \\
\label{a.9b}
&& \left[ \hbar \Omega_{\alpha} - H_i\right] Y^{(\alpha)}_i = -
 \sqrt{2S} \sum_{p=1}^{N_h}\sum_{h=1}^{N_d}
 J_{p\downarrow,h\uparrow}^*(i) X^{(\alpha)}_{hp}
\end{eqnarray}

These are the RPA equations and they can be recast in the customary
standard eigenvalue RPA equation form\cite{Ripka}
\begin{equation}
\label{5.2}
\hbar \Omega \left(
\begin{array}[c]{c}
{\bf X} \\ {- \bf Y} \\
\end{array}
\right) = \left(
\begin{array}[c]{cc}
{\bf E} & {\bf J}\\ {\bf J}^* & {\bf H} \\
\end{array}
\right) \cdot \left(
\begin{array}[c]{c}
{\bf X} \\ { \bf Y} \\
\end{array}
\right)
\end{equation}
where the vectors ${\bf X}$ and ${\bf Y}$ contain all the unknowns
$X_{hp}$ and $Y_i$, and the matrices ${\bf E}, {\bf H}$ and ${\bf J}$
have the elements ${\bf E}_{hp,h'p'} = -\delta_{hh'} \delta_{pp'}
(E_{h\uparrow} - E_{p\downarrow})$, ${\bf H}_{i,i'} = -\delta_{i,i'}
H_i$ and ${\bf J}_{hp, i} = \sqrt{2S} J_{p\downarrow,h\uparrow}(i)$.

 The dimension of the RPA matrix (\ref{5.2}), and therefore the number
of normal modes, is $N_d + N_h\times N_d$. Of these solutions, $N_d$
are proper spin-wave collective modes, while the rest of $N_h\times
N_d$ are spin-flip processes associated with particle-hole
excitations.  If there were no interactions (J=0), the eigenenergies
of these spin-flip processes would be $H_i$ for the lowering of the Mn
spin $i$, respectively $E_{h\uparrow} - E_{p\downarrow}$ for a hole
spin-flip [see the left-hand side of
Eqs. (\ref{a.9}),(\ref{a.9b})]. Interactions renormalize these values
[as described by the right-hand-side of
Eqs. (\ref{a.9}),(\ref{a.9b})], but one still expects $N_d$ spin-wave
collective solutions at low energies, coming from the Mn spin
lowering, and $N_h\times N_d$ spin-flip particle-hole excitations at
energies comparable or larger than the spin-flip gap $\Delta=
E_{1\uparrow} - E_{N_h\downarrow}$.

We now comment on the sign of the RPA spectrum frequencies
$\hbar\Omega_{\alpha}$. Since $H_i > 0$ [see Eq. (\ref{29})] and
$E_{h\uparrow} - E_{p\downarrow}> 0$ (by definition of the HF
ground-state), it is apparent from Eqs. (\ref{a.9}),(\ref{a.9b}) that
$N_d$ spin-wave solutions will have positive energies
$\hbar\Omega_{\alpha}$, while the $N_h\times N_d$ spin-flip
particle-hole excitations will have negative energies. This {\em does
not} imply that the system is unstable, but simply that we have not
chosen the proper spin-flip creation operators. Instead of the choice
of Eq. (\ref{a.8}), we can also denote
\begin{equation}
\label{a.10}
P_{\alpha}=Q^{\dagger}_{\alpha}= \sum_{h=1}^{N_d}\sum_{p=1}^{N_h}
X_{hp}^{(\alpha)} b_{hp} + \sum_{i}^{} Y_i^{(\alpha)} B^{\dagger}_i
\end{equation}
In terms of these new operators, we have now $[{\cal H}_{RPA},
P_{\alpha}]= \hbar \Omega_{\alpha} P_{\alpha}$, i.e.
\begin{equation}
\label{a.11}
{\cal H}_{RPA} = \sum_{\alpha}^{}(-\hbar\Omega_{\alpha})
P^{\dagger}_{\alpha} P_{\alpha}
\end{equation}
In other words, a negative solution for $\hbar \Omega$ simply means
that we chose the corresponding annihilation operator instead of the
creation operator when we wrote Eq. (\ref{a.8}). For spin-flip
processes, it is obvious that this is the case from the dependence of
$Q^{\dagger}_{\alpha}$ on $b_{hp}$, instead of $b_{hp}^{\dagger}$.

An unstable mean-field ground-state is signaled by {\em complex}
values of the spectrum frequencies $\hbar
\Omega_{\alpha}$.\cite{Rowe,Ripka} Excited states about the mean-field
ground-state $|\Psi\rangle$ are of the general form $|\Phi\rangle = [1
+ \sum_{}^{} c_{\alpha} (Q^{\dagger}_{\alpha})^{n_{\alpha}} ]
|\Psi\rangle$. Within RPA the dynamics of such states is dictated by
${\cal H}_{RPA}$, leading to a time dependence:
$$ |\Phi(t)\rangle = e^{-i E_{HF}t/\hbar} [ 1 + \sum_{\alpha}^{}
  c_{\alpha} e^{ - i n_{\alpha} \Omega_{\alpha} t}
  (Q^{\dagger}_{\alpha})^{n_{\alpha}}] |\Psi\rangle,
$$ where $E_{HF}=\langle \Psi | {\cal H} | \Psi \rangle$ is the
Hartree-Fock energy of the system.  If any of the frequencies
$\Omega_{\alpha}$ has a non-trivial imaginary part, it follows that
the expectation value of any operator $\langle \Phi(t) | ... | \Phi(t)
\rangle$ will move in time exponentially away from its mean-field
ground-state expectation value $\langle\Psi | ... | \Psi \rangle$,
i.e. the mean-field is unstable to small perturbations.

The advantage of the standard RPA approach is that by solving the
eigenvalue problem (\ref{5.2}), one finds the normal mode spectrum
$\hbar \Omega$ and the spatial distribution of the spin-waves (related
to the $Y_i$ coefficients) at once. The obvious disadvantage is of a
numerical nature: for a disordered system the RPA matrix must be
diagonalized numerically. The size of the matrix is $n
=(1+N_h)N_d$. The typical sizes we consider are systems with around
$N_d =500$ Mn spins, and $N_h=50=10\% N_d$ holes (although
concentrations up to $30\%$ might have to be considered, depending on
the doping $x$), leading to RPA matrices of typical size $n >
25,000$. As a result, one has to either consider much smaller systems
(in which case, finite size effects may be overwhelmingly important),
or to try sparse matrix techniques to obtain the first few low-energy
modes of the RPA matrix (although the off-diagonal ${\bf J}$ matrix is
not necessarily sparse).

There is, however, an alternative way of finding the collective mode
spectrum and spatial distributions. We can directly solve for the
$X_{hp}$ coefficients from Eq. (\ref{a.9}) and rewrite
Eq. (\ref{a.9b}) as
\begin{equation}
\label{5.3}
\sum_{i'}^{} M_{ij}(\omega) Y_j(\omega) = \hbar \omega Y_i(\omega)
\end{equation}
where
\begin{equation}
\label{5.4}
M_{ij}(\omega) = \delta_{ij} H_i- 2S \sum_{p=1}^{N_h} \sum_{h=1}^{N_d}
{ J^*_{p\downarrow,h\uparrow}(i)J_{p\downarrow,h\uparrow}(j) \over
\hbar \omega + E_{h\uparrow} - E_{p\downarrow} +i\eta }
\end{equation}
The advantage of this formulation is that one has to deal with much
smaller matrices (the dimension of the ${\bf M}$ matrix is $N_d=500$).
Eq. (\ref{5.3}) tells us that for frequencies $\Omega$ belonging to
the spin-wave spectrum, the matrix ${\bf M}(\Omega)$ has at least one
eigenvector corresponding to an eigenvalue
$\lambda(\Omega_{\alpha})=\hbar\Omega_{\alpha}$ (if the mode is
degenerate, there are several such eigenvectors). The corresponding
eigenvectors give us the desired spatial mode distribution
$Y_i^{(\alpha)}=Y_i(\Omega_{\alpha})$. Thus, the problem is reduced to
sweeping the range of frequencies of interest (for low-energy
collective excitations, this is usually a fairly small range of
frequencies near $\omega=0$), and monitoring the eigenvalues of the
${\bf M}(\omega)$ matrix.  The dependence on $\omega$ of the matrix
elements $M_{ii'}(\omega)$, and therefore of the eigenvalues
$\lambda(\omega)$ is monotonic for small $\omega\ll \Delta$ [see
Eq. (\ref{5.4})]. As a result, each equation
$\lambda(\omega)=\hbar\omega$ has a single solution, and we expect to
find $N_d$ collective mode eigenenergies, one for each eigenvalue. The
monotonic dependence of $\lambda(\omega)$ on $\omega$ also simplifies
the search for the collective spectrum, since it is enough to evaluate
the eigenvalues $\lambda(\omega)$ in the range of interest on a grid
of step $\delta\omega$, and use linear interpolation to find the
solutions of the equations $\lambda(\omega)=\hbar\omega$. As we show
in the following by comparison with standard RPA (for small system
sizes) we can very easily achieve relative errors less than $10^{-5}$.

The RPA spectrum has a second type of solutions, the spin-flip
particle-hole continuum. These solutions are at frequencies of the
order of the gap between the last occupied and first empty HF states
$\hbar \omega \sim \Delta$, and one expects $N_h\times N_d$ such
solutions. In principle, one can use the same method to find these
energies as well, although the fact that they extend over a larger
range of frequencies and that the $\omega$ dependence is very strong
makes the search more difficult.  Typically, one is only interested in
the lower and upper limits of the spin-flip continuum, which can be
found easily with the method just described.

Finally, we would like to mention that it is always possible to
reformulate the standard RPA equation (\ref{5.2}) in a
``self-consistent'' form of the type shown in Eq. (\ref{5.3}), with a
matrix ${\bf M}(\omega)$ of much smaller dimension than the RPA
matrix. For instance, for Hamiltonians describing interacting electron
systems, the RPA matrix has a dimension $\sim N_oN_e$, where $N_o$ is
the number of occupied, and $N_e$ is the number of empty orbitals in
the mean-field ground-state.\cite{Ripka} For a system with a finite
concentration of electrons $x$ on a lattice of linear dimension $N$ in
a $D$-dimensional space, we have $N_o \sim x N^D$ and $N_e \sim
(1-x)N^D$, and the RPA matrix scales as $x(1-x)N^{2D}$.  For such
systems, one can always find an equivalent reformulation of the RPA
equation in terms of a ${\bf M}(\omega)$ matrix of size $\sim
N^D$.\cite{thesis} This allows numerical handling of considerably
larger systems, without having to resort to sparse matrix techniques,
which often have issues related to instability.

\section{Implementation}
\label{sec4}

In this section we show, by direct comparison against known
calculations, as well as by direct comparison against solving the RPA
matrix equation, that the formulation of Eqs. (\ref{5.3}), (\ref{5.4})
gives correct and numerically very accurate results. In the second
part of the section we describe in more detail the numerical
implementation as well as efficiency of our method.

\subsection{Analytical solution}

We first apply our approach by solving Eqs. (\ref{5.3}), (\ref{5.4})
analytically for a simplified case. We assume that the Mn spins are
arranged on an ordered superlattice, instead of having random
positions. Then, the charge-carrier part of the mean-field Hamiltonian
is easily diagonalized in the $\vec{k}$-space:
\begin{equation}
\label{6.1}
{\cal H}_{cc}^{HF} = \sum_{\vec{k}\sigma}^{} E_{\vec{k}\sigma}
c^{\dagger}_{\vec{k}\sigma} c_{\vec{k}\sigma}
\end{equation}
where
\begin{equation}
\label{6.2}
E_{\vec{k}\sigma} = \epsilon_{\vec{k}} + { J_0 S\over 2} \sigma
\end{equation}
Here, $\epsilon_{\vec{k}} = \sum_{\vec{\delta}\ne 0}^{}
t_{\vec{\delta}} \exp{(i\vec{k}\cdot\vec{\delta})}$ is the kinetic
energy of the non-interacting electrons, where $\vec{\delta}$ indexes
all the neighboring sites and $t_{\vec{\delta}}= t_{ij}$ for which
$|\vec{R}_i - \vec{R}_j| = | \vec{\delta}|$. Also, $J_0 =
\sum_{\vec{\delta}} J_{\vec{\delta}}$, where again $J_{\vec{\delta}}=
J_{ij}$ for which $|\vec{R}_i - \vec{R}_j| = | \vec{\delta}|$.  In the
HF ground state, the states $(\vec{k}, \downarrow)$ given by the
condition $N_h =\sum_{|\vec{k}| < k_F} 1 $ are occupied.  Since the
$N_h$ electrons are equally distributed among the $N_d$ Mn sites, and
are fully polarized, the z-component of the total spin created by
electrons at each site, within HFA, is $ s_h(j) = - { 1 \over 2} {N_h
\over N_d} = - { p / 2}$, and therefore we find $H_i = J_0 p / 2$ [see
Eq. (\ref{29})].

Using the fact that the solutions of Eq. (\ref{5.3}) must also have
plane-wave structure $Y_i \sim \exp{(i\vec{Q}\cdot\vec{R}_i)}$, it
follows that the index $\alpha\rightarrow \vec{Q}$ becomes a
wave-vector, as expected, and we can reduce Eqs. (\ref{5.3}) and
(\ref{5.4}) to a self-consistent equation for each collective mode
$\hbar \omega(\vec{Q})$:
\begin{equation}
\label{55}
\hbar \omega - {p J_0\over 2} = - {S|J(\vec{Q})|^2 \over 2
 N_d}\sum_{\vec{k}}^{} { f(E_{\vec{k},\downarrow}) \over \hbar \omega
 + \epsilon_{\vec{k}-\vec{Q}} - \epsilon_{\vec{k}} + J_0 S}
\end{equation}
Here, $J(\vec{Q}) = \sum_{\vec{\delta}}
\exp{(i\vec{Q}\cdot\vec{\delta})} J_{\vec{\delta}}$. The occupation
number $f(E_{\vec{k},\downarrow})$ obviously comes from the sum over
occupied states we had in Eq. (\ref{5.4}). For a finite-size ordered
lattice, one expects considerable degeneracy for each mode
$E_{\vec{k}\sigma}$, due to invariance to various symmetry
transformations of the $\vec{k}$-vector. In most cases, the number of
charge carriers is such that the last orbital (of degeneracy $G$) is
only partially occupied by $g$ charge carriers. In this case, one must
obviously choose $f= g/G$ for all these orbitals, and $f=1$ for lower,
fully occupied orbitals, and $f=0$ for higher, empty
orbitals. Otherwise, the translational invariance is broken and the
collective spectrum is not indexed by a wave-vector.

The sum in Eq. (\ref{55}) can be performed if we assume that the
dispersion at the bottom of the band is of quadratic form $
\epsilon_{\vec{k}} = \hbar^2 k^2 / (2m)$ (this is a reasonable
approximation since we are interested in low filling fraction
$p$). Then, the sum over occupied states can be transformed to an
integral which is straightforward to evaluate, leading to the
solution:
\begin{equation}
\label{56}
\hbar\omega= {p J_0\over 2} + |J(\vec{Q})|^2 { 3 p S\over 4 v_F} { 1
\over Q } f(\omega, {Q})
\end{equation}
Here, $v_F = \hbar k_F/m$ is the Fermi velocity, where the Fermi
vector is given by
$$ N_h = p N_d = { V \over (2\pi)^3} { 4 \pi k_F^3\over 3}.
$$ The function $f(\omega, Q)$ is given by
$$ f(\omega, Q) = - { { \hbar\omega + \epsilon_{Q} + J_0S} \over v_F
Q}
$$
$$ +{1 \over 2} \left[ 1 - \left( {{ \hbar\omega + \epsilon_{Q} +
J_0S} \over v_F Q}\right)^2\right] \ln{ { \hbar\omega + \epsilon_{Q} +
J_0S- v_F Q} \over { \hbar\omega + \epsilon_{Q} + J_0S+ v_F Q}}
$$ where $\epsilon_{Q}= \hbar^2 Q^2/2m$.

  For comparison purposes, we will assume a local AFM interaction
$J_{ij} \rightarrow c J_{pd} \delta_{ij}$, leading to $J(\vec{Q}) =
J_0= c J_{pd}$. In this case, the Hamiltonian becomes identical to the
Hamiltonian used by Konig. {\em et. al} in Ref. \onlinecite{MacD},
provided that the effective mass $m$ in the dispersion relation is
assumed to be the band effective mass. We can easily solve
Eq. (\ref{56}) in the asymptotic limit $Q\rightarrow 0$, to find
\begin{equation}
\label{6.6}
\hbar \omega(\vec{Q}) = { \hbar^2 Q^2 \over 2m} {1\over 2S/p -1 }
\left[ 1 - { 4 \over 5} { t \over J_0S} \left( { p \over 6
\pi^2}\right)^{2/3}\right]
\end{equation}
This is indeed identical with the asymptotic long-wavelength spin-wave
spectrum obtained in Ref. \onlinecite{MacD}, and has the typical
quadratic dependence of spin-waves in conventional ferromagnetic
systems.

\subsection{Direct comparison between the two formulations}

In this section we briefly illustrate the accuracy and speed of our
formulation of the RPA problem [Eqs. (\ref{5.3}) and (\ref{5.4})], in
comparison with the standard RPA approach [Eq. (\ref{5.2})].  We use a
rather small system, with $N_d = 80$ Mn spins randomly distributed on
a FCC lattice with linear size $N=10$, corresponding to a Mn
concentration $x=N_d/4N^3=0.02$. The number of charge carriers is
$N_h=pN_d=8$, and the other parameters are as defined in Section
\ref{sec2}.  Thus, the standard RPA involves the diagonalization of a
non-hermitian matrix of dimension $720$.

\begin{figure}
\centering
\includegraphics[angle=270,width=\FigWidth]{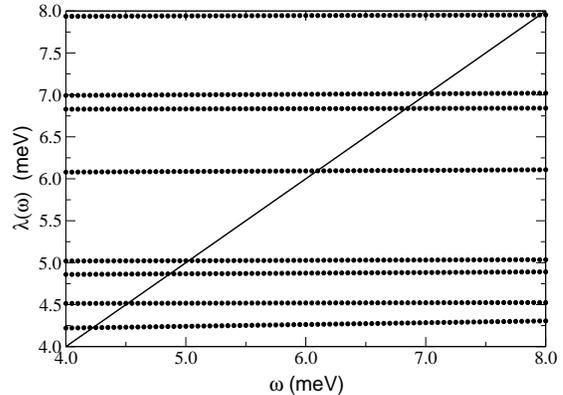}
\caption{\label{fig1} The largest eight eigenvalues $\lambda(\omega)$
of the ${\bf M}(\omega)$ matrix (circles), evaluated on a grid with a
step $\delta\omega=0.05$ meV. The full line is $\hbar\omega$. The
spin-wave spectrum is given by the condition
$\lambda\omega=\hbar\omega$.  }
\end{figure}

\begin{figure}[b]
\centering
\includegraphics[angle=270,width=\FigWidth]{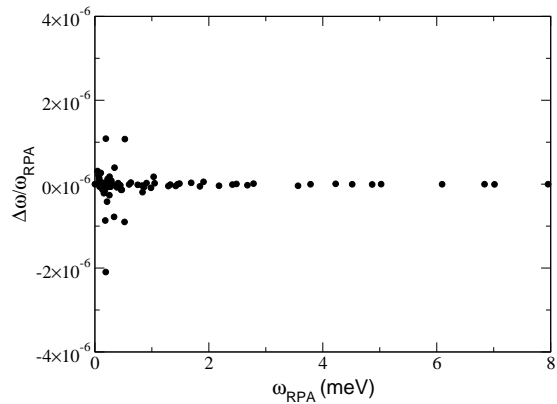}
\caption{\label{fig2} Accuracy $\Delta\omega/\omega_{RPA}$ for all
$N_d=80$ spin-wave frequencies $\omega_{RPA}$, between the exact RPA
(standard) values and values obtained in our formulation using a grid
with $\delta\omega=0.05$ [see Fig. \ref{fig1}].}
\end{figure}

In Fig. \ref{fig1} we show the $\omega$-dependence of the largest
eight eigenvalues $\lambda(\omega)$ of the ${\bf M}$ matrix (circles),
evaluated on a grid with a step $\delta\omega=0.05$ meV. [For real
$\omega$ and $\eta=0$, the matrix ${\bf M}(\omega)$ is hermitian and
all eigenvalues $\lambda(\omega)$ are real, see Eq. (\ref{5.4})].  The
full line is $\hbar\omega$, and the spin-wave spectrum is given by the
condition $\lambda(\omega)=\hbar\omega$. One can see that the
eigenvalues $\lambda(\omega)$ have monotonically increasing dependence
on $\omega$ for small $\omega \ll \Delta\approx 50$ meV, and therefore
the equation $\lambda(\omega)= \hbar\omega$ can have at most one
solution for each eigenvalue, for small $\omega$. If each eigenvalue
yields a solution, we have found all $N_d$ spin-wave frequencies. If
one or more eigenvalues do not intersect the $ \hbar\omega$ line, this
means that the mean-field ground-state is unstable ( there are {\em
complex} spin-wave frequencies). For all the cases and parameters we
investigate, we found that the fully-polarized ground-state is stable
for our model.

\begin{figure}
\centering
\includegraphics[angle=270,width=\FigWidth]{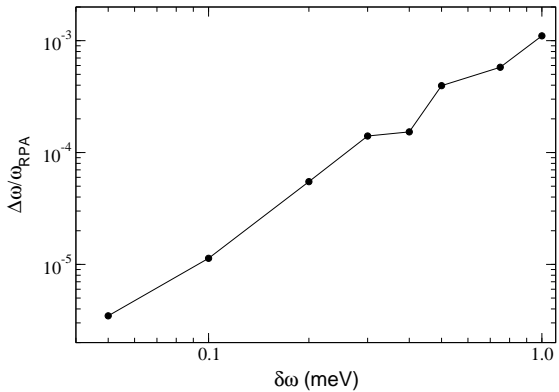}
\caption{\label{fig3} Largest relative error
$\Delta\omega/\omega_{RPA}$ among $5N_d=400$ spin-wave frequencies
$\omega_{RPA}$, between the exact RPA (standard) values and values
obtained in our formulation using a grid with
$\delta\omega=0.05,0.1,0.2,0.3,0.4, 0.5, 0.75$ and 1 meV. The line is
drawn to guide the eye. Even for a very large grid, the relative error
is still reasonably small.}
\end{figure}

To find the spectrum frequencies $\hbar\Omega_{\alpha}$, we use linear
interpolation over the interval $\delta\omega$ where each
$\lambda(\omega)$ curve intersects the $\hbar\omega$ curve.  This
method avoids the need of evaluating the eigenvalues at too many
$\omega$ points (the most time-consuming part of the computation is
the evaluation of the matrix elements of ${\bf M}$).  The high degree
of accuracy obtained with this method is demonstrated in
Fig. \ref{fig2}, where we compare the exact RPA spectrum
$\omega_{RPA}$ obtained through direct diagonalization of the RPA
matrix [see Eq. (\ref{5.2})] with the the values ${\tilde
\omega}_{RPA}$ obtained through our formulation. In fact, we plot the
relative error $\Delta\omega/\omega_{RPA} = ({\tilde
\omega}_{RPA}-\omega_{RPA}) / \omega_{RPA}$ for each spectrum
frequency $\omega_{RPA}$. As one can see from Fig. \ref{fig2}, the
largest error is of the order $2 \cdot 10^{-6}$, for the grid
$\delta\omega=0.05$.

This suggests that one could use a much larger grid $\delta\omega$ and
still have a good accuracy. Indeed, we have computed the spectrum
through both methods for five different realization of disorder, using
various grid values $\delta\omega$ in our formulation, and selected
the largest relative error for each case (from a total of $5N_d=400$
values). The results are shown in Fig. \ref{fig3}. The error is found
to scale as the square of the grid size. Even with a very large grid
$\delta\omega=1$ meV, (which implies the evaluation of the eigenvalues
$\lambda(\omega)$ in only very few points), we still get a relative
error smaller than $10^{-3}$. Thus, one can use the grid step
$\delta\omega$ to optimize and considerably speed up the
computation. However, as the number of spins (and eigenvalues) $N_d$
increases, one must take into consideration other complications, as
discussed below.

\begin{figure}
\centering
\includegraphics[angle=270,width=\FigWidth]{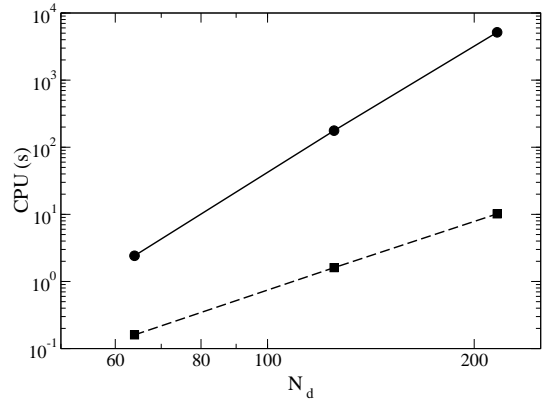}
\caption{\label{fig4} CPU time for the two RPA methods, for systems
with $N_d=64, 125$ and 216 spins. The standard RPA (full line) is
significantly more time consuming than our formulation (dashed line).}
\end{figure}

In Fig. \ref{fig4} we compare CPU times for finding the RPA spin-wave
spectrum using the standard vs. our RPA formulations. All simulations
were done on the same processor. We used a small grid
$\delta\omega=0.05$ meV for our approach, and verified that all
relative errors were less than $10^{-5}$. We use systems with $N_d=64,
125$ and 216 spins, randomly distributed on FCC lattices of linear
size $N=12, 15$ and 18 ($x=0.0092, p=10\%$).  As expected, the
computational time depends in a power-law manner on the size $N_d$,
with exponents 6.4 and 3.2 respectively for the standard and current
approach. Clearly, our formulation can be successfully used for much
larger sizes than those afforded by the standard approach.

\begin{figure}
\centering
\includegraphics[angle=270,width=\FigWidth]{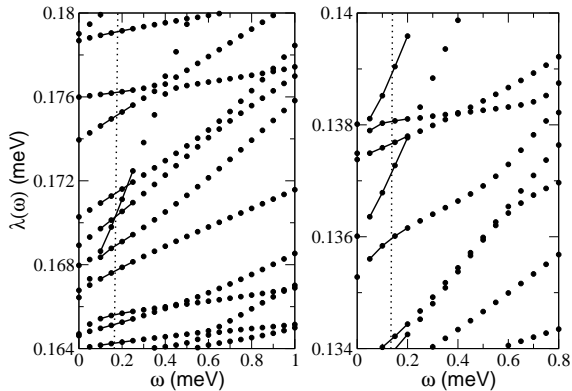}
\caption{\label{fig5} Eigenvalues $\lambda(\omega)$ for a system with
$N_d=512$ random spins, evaluated for a grid
$\delta\omega$=0.05meV. Crossing and anti-crossings of neighboring
eigenvalues is apparent. The dotted line is $\hbar\omega$. Since we
are interested in the solutions of the equations $\lambda(\omega)=
\hbar\omega$, it is clear that too large a grid $\delta\omega$ may
lead to considerable errors .}
\end{figure}

One important aspect to keep in mind is that as the system size $N_d$
increases, neighboring eigenvalues become more closely spaced, and
accidental crossings and anti-crossings appear.  Some typical examples
are shown in Fig. \ref{fig5}, for a disordered system with $N_d=512$
spins. Tracking the crossing of the eigenvalues is essential, since
such events lead to a change in the indexing of the eigenvalues from
one grid point to the next. We found that requiring continuity in
first and second derivatives allows a unique identification of each
continuous eigenvalue from one grid point to the next (3 steps are
shown for each eigenvalue in Fig. \ref{fig4}). In fact, the only
relevant crossings are the ones that take place within the step
$\delta\omega$ where the eigenvalues intersects the $\hbar\omega$ line
(shown as a dotted line in Fig. \ref{fig4}). Such an example is shown
in the left panel of Fig. \ref{fig4}. For the largest system size
investigated, of $N_d=512$ randomly distributed spins, we find that
only 4.5\% of the eigenvalues have such relevant crossings within
$\delta\omega$ of $\lambda(\omega)=\hbar\omega$. The percentage
decreases with decreasing $N_d$, to 1.1\% and 0.3\% for $N_d=$ 216 and
125, respectively. This percentage also depends on the grid step
$\delta\omega$; for increasing $\delta\omega$ the identification of
crossings and anti-crossings becomes more difficult, leading to
possibly large errors. One can optimize the choice of the grid step
$\delta\omega$ by starting with a larger value. The well separated
eigenvalues (such as the ones depicted in Fig. \ref{fig1}) will
provide unique identification and very accurate values for their
corresponding spin-wave spectra values. However, where considerable
mixing and therefore non-linear variation of the eigenvalues
$\lambda(\omega)$ is apparent, a finer mesh is necessary in order to
correctly characterize their variation near $\hbar\omega$.  In all our
simulations, we use the grid $\delta\omega=0.05$ meV, which allows for
comfortable tracking of each eigenvalue and also is sufficiently small
to allow us to approximate the variation of $\lambda(\omega)$ as being
linear within each $\delta\omega$ step.

These comparisons clearly demonstrate the accuracy and speed of our
formulation of RPA, as compared to the standard RPA. The biggest
advantage, though, is that it can easily be applied to systems with
large sizes, for which standard RPA is numerically cumbersome.

\section{Spin-wave spectra of DMS}
\label{sec5}

In this section, we use the RPA method described above to study the
density of states (DOS), as well as nature (extended or localized) of
the spin-wave spectrum of (III,Mn)V DMS described by the model
introduced in Section \ref{sec2}.

\begin{figure}
\centering
\includegraphics[angle=270,width=\FigWidth]{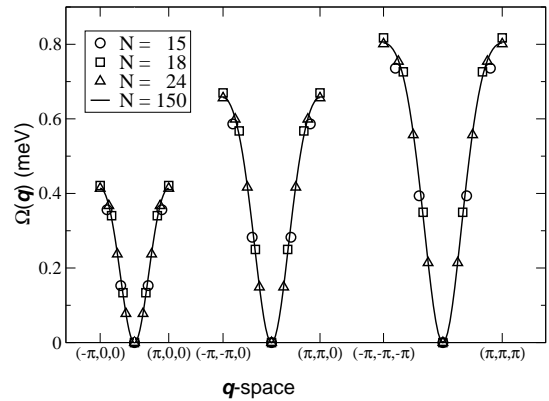}
\caption{\label{fig6} Spin-wave dispersion $\Omega(\vec{q})$ for an
ordered, simple cubic superlattice arrangement of Mn spins inside the
host semiconductor. Lattices of linear size $N=15,18,24$ and 150, with
a total of respectively $125, 216, 512$ and $125000$ Mn spins are
considered. This corresponds to $x=0.0092$ and $p=10\%$. Dispersion is
plotted along three linear cuts in the Brillouin zone: $(-\pi,0,0)
\rightarrow (\pi,0,0)$, $(-\pi,-\pi,0) \rightarrow (\pi,\pi,0)$ and
$(-\pi,-\pi,-\pi) \rightarrow (\pi,\pi,\pi)$.  }
\end{figure}

\subsection{Spin-wave density of states}

We study the spin-wave density of states for three types of
arrangements of the Mn spins inside the host semiconductor. All
samples studied correspond to $x=0.0092$ and $p=10\%$. Other
parameters are as specified in Section \ref{sec2}.

	 First, we consider fully ordered systems, in which the Mn
spins are arranged on a simple cubic superlattice. For $x=0.0092$, the
superlattice constant is $a_L=3a$. In this case, we can study the
spin-wave dispersion and density of states for very large sizes, since
we can use directly Eq.  (\ref{56}) to compute the spin-wave frequency
$\hbar \omega(\vec{q})$ for each wave-vector $\vec{q}$ inside the
Brillouin zone. Spin-wave dispersion obtained along three cuts in the
Brillouin zone is shown in Fig. \ref{fig6}. The dispersion has
quadratic behavior near the center of the Brillouin zone, as expected
from the discussion for an ordered case provided above. The
finite-size effects are reasonably small. For the small sizes we used
both Eq. (\ref{56}) and our method to compute the dispersion. The
results of the two agree with a relative error of less than $10^{-6}$.
	
One important aspect to notice is the small range of the spin-wave
spectrum, as compared to the AFM exchange $J=15$ meV. This is a
consequence of the fact that the Mn spins do not interact directly
with one another. Instead, their interaction is mediated by the rather
small concentration of charge carriers present.

\begin{figure}
\centering
\includegraphics[width=\FigWidth]{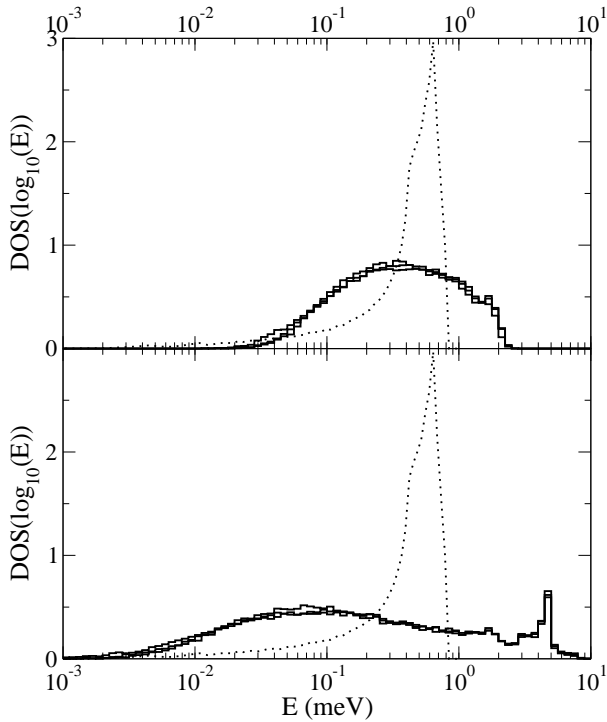}
\caption{\label{fig7} Upper panel: average density of states
$\rho(\log_{10}E)$ {\em on a logarithmic scale} for systems with
$N_d=125, 216$ and 512 Mn spins in moderately disordered
configurations (full lines). Lower panel: average density of states
for systems with $N_d=125, 216$ and 512 Mn spins in strongly
disordered configurations (full lines). The dotted line is the
spin-wave density of states of a DMS with fully ordered (superlattice)
configuration of Mn spins. All samples correspond to $x=0.00924$,
$p=10\%$. }
\end{figure}

We compute the density of states (DOS) $\rho(E)$ associated with the
spin-wave dispersion for the superlattice case employing the standard
method of dividing the Brillouin zone into tetrahedra and linearizing
the dispersion (Ref. \onlinecite{tetrah}).  The DOS obtained for the
lattice with $N=150$, $N_d=125,000$ is shown as a dotted line in
Fig. \ref{fig7}. We use a logarithmic scale for the energy, and
normalize the density of states such that
$$ \int_{-\infty}^{\infty} dx \rho(x) = 1
$$ where $x = \log_{10}(E)$. This convention will be maintained
throughout the rest of the paper.

In the upper panel of Fig. \ref{fig7} we also plot the densities of
states obtained for moderately disordered configurations with
$N_d=125,215$ and $512$ Mn spins (full lines). These are
configurations in which we place the Mn spins randomly on the FCC Ga
sublattice of the host semiconductor, subject to the constraint that
the distance between any two spins is {\em larger than $2a$}. [In the
ordered cubic superlattice, nearest neighbor spin separation is
$a_L=3a$].  This moderate amount of disorder breaks translational
invariance, and the wave-vectors are no longer good quantum
numbers. Also, the large degeneracies of the superlattice spectrum are
lifted. We computed the spin-wave spectrum for $200$ realizations of
the disorder with $N_d=125$ spins, 100 realizations with $N_d=216$
spins and 50 realizations with $N_d=512$ spins. Thus, we have a total
of over 21,000 spin-wave energies for each size, and statistics can be
comfortably carried out. In particular, from Fig.  \ref{fig7} we see
that the DOS histograms corresponding to the three sizes are smooth
functions, i.e. the average over disorder is well accounted for. Also,
the curves are practically indistinguishable from one another,
implying that finite size effects are negligible.

In the lower panel of Fig. \ref{fig7} we plot the DOS corresponding to
fully disordered configurations with $N_d=125,215$ and $512$ Mn spins
(full lines). These are configurations in which we place the Mn spins
randomly on the FCC Ga sublattice of the host semiconductor, with no
restrictions (except that each Mn occupies a different site). The
number of samples analyzed is the same as in the previous case. Again,
the curves show smooth behavior and no finite-size effects. The origin
of the peak appearing near $\omega \sim 5~$meV will be discussed
later.

Before continuing the analysis, we must also point out that the
Goldstone modes have been left out in the DOS shown in
Fig. \ref{fig7}. The RPA system {\em always} has one solution of
energy $\omega=0$ (numerically, we find its magnitude to be less that
$10^{-14}$), corresponding to $Y_i=1/\sqrt{N_d}$. This is the
Goldstone mode, describing the same overall rotation of all the
spins. These Goldstone modes would appear in the DOS as a
$\delta(\omega)$ function at $\omega=0$ ($\log_{10}E= - \infty$).

The effect of disorder in the positions of the Mn is reflected in the
considerable widening of the DOS (on a logarithmic scale), and
rounding of the van Hove singularities of the superlattice DOS, as the
amount of disorder increases.  More significantly, however, is the
{\em substantial} enhancement of the DOS at {\em low energies}.  This
behavior is in agreement with the general expectation of the effects
of disorder on an energy-band dispersion.

A question of considerable interest concerns the {\em nature} of the
 spin-wave excitations. We know that the ordered superlattice has
 extended, plane-wave type spin-waves. The general expectation is that
 disorder will lead to localization, and thus one expects the
 appearance of localized spin-waves as the disorder increases. One way
 to characterize the nature of the spin-waves is to compute their
 Inverse Participation Ratio (IPR), defined as:
\begin{equation}
\label{60}
\rm{IPR}_{\alpha} = {{\sum_{i=1}^{N_d} \left(Y_i^{(\alpha)}\right)^4}
\over {\left(\sum_{i=1}^{N_d} \left(Y_i^{(\alpha)}\right)^2\right)^2}}
\end{equation}
For an extended mode $\alpha$ (plane-wave, for instance), we expect
that all $Y_i^{(\alpha)}$ coefficients are roughly of equal magnitude,
since all spins are expected to participate equally in the
spin-wave. Then, it follows that for an {\em extended mode $\alpha$},
IPR$_{\alpha} \sim 1/N_d$, i.e. it is inversely proportional to the
size of the system. On the other hand, in a localized mode $\alpha$,
only the spins within the localization volume have non-vanishing
values for $Y_i^{(\alpha)}$. Thus, it follows that for a {\em
localized mode $\alpha$}, IPR$_{\alpha}$ is independent of
$N_d$. Instead, its value is given by the inverse number of Mn spins
participating in the localized mode.

\begin{figure}
\centering
\includegraphics[width=\FigWidth]{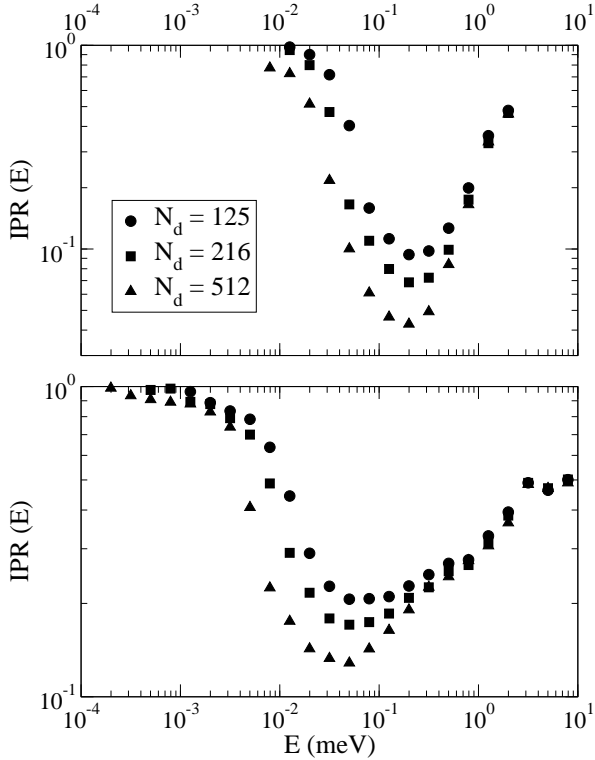}
\caption{\label{fig8} Upper panel: average IPR(E) for systems with
$N_d=125, 216$ and 512 Mn spins in moderately disordered
configurations (circles, square and triangles, respectively). Lower
panel: same for strongly disordered configurations. All samples
correspond to $x=0.00924$, $p=10\%$. }
\end{figure}

We compute the IPR of the spin-waves in the following manner. As we
sweep the frequencies $\omega$ of interest and compute the eigenvalues
$\lambda(\omega)$ of the ${\bf M}(\omega)$ matrix on the grid
$\delta\omega$, we also compute the corresponding IPR of the
eigenvectors ${\bf Y(\omega)}$ (which are normalized to unity). The
IPR is a continuous function of $\omega$, and we use linear
interpolation to find its value at the spin-wave frequencies
$\Omega_{\alpha}$ of interest. Whenever eigenvalues $\lambda(\omega)$
cross, the IPRs are no longer well defined: one can choose any linear
combination of the eigenvectors of the degenerate modes, which would
lead to different values for the corresponding IPRs.  We have checked
that these accidental crossings are rather rare (below a few percent)
and equally distributed over the entire energy scale. If we exclude
all these degenerate cases, we get a DOS which is virtually
indistinguishable from the one obtained when these modes are kept.
More importantly, we find that modes which cross predominantly have
the same nature (either localized or extended). As a result, the
ensemble averaged IPR is not sensitive to the precise treatment of
these cases.

In Fig. \ref{fig8} we plot the (geometric) average IPR(E) for the
moderate (upper panel) and strong (lower panel) disorder
configurations.  Again, the Goldstone modes are not shown. We use the
geometric mean (i.e., arithmetic mean of the $\log$ values) in order
to insure proper weight for the extended modes, with low IPR.  For
moderate disorder configurations, we see that the spin-wave modes at
high energies $E > 1$ meV are localized: the values corresponding to
the three different sizes collapse on top of one another. This is
expected, given the fact that the upper band-edge of the superlattice
spectra is just below $E = 1$ meV (see Fig. \ref{fig7}). Thus, states
of higher energy have been split off the band by disorder, and are
expected to be localized. The low-energy part of the band also appears
to be localized: while the three curves do not quite meet, the value
of the IPR is just below unity, showing that these spin-waves have
considerable weight on only one spin (i.e., they correspond to a
individual spin-flip). However, for the central part of the spectrum,
the spin-waves are extended, with the IPRs for the three sizes clearly
distinct and decreasing with increasing $N_d$.

\begin{figure}
\centering
\includegraphics[angle=270,width=\FigWidth]{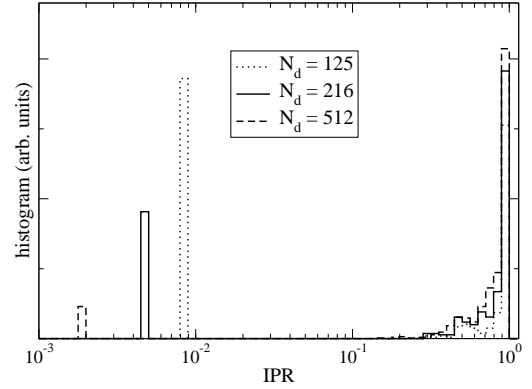}
\caption{\label{fig9} Histogram of the IPR values for all spin-waves
(spin flips) of energy $E < 5\cdot 10^{-3})$, for $N_d = 125, 216$ and
512.  Results are shown only for strong disorder configurations.
Except for the low-lying Goldstone modes, the histograms are
identical, suggesting only localized modes at these energies.  }
\end{figure}

For the strong disorder configurations, this tendency is even more
apparent. The high- and low-energy regions, $E > $ 1 meV, respectively
$E < 5\cdot 10^{-3}$ meV, contain localized spin-waves: in both
limits, the IPR curves collapse on top of one another. The low-energy
localized modes are individual spin-flips. The associated $Y_i$ are
vanishingly small at all sites except one, leading to IPR $\approx
1$. These sites are always situated far apart from all other Mn spins,
and the probability to be visited by charge carriers is exponentially
small (tailing from far occupied regions). As a result, the Mn spins
at these sites are virtually isolated, and their spin excitations are
individual spin-flips. The energy for such a spin-flip is equal to
$H_i$, if one neglects small corrections due to the extremely weak
interactions [see Eq. (\ref{a.9b}) and following discussion].

Histograms of the IPR of all the modes with energies below $E < 5\cdot
10^{-3})$ are shown in Fig. \ref{fig9}. Here, we also show the
Goldstone modes, which have zero energy, and $IPR=1/N_d$. The
histograms have been scaled by the total number of modes for all
disorder realizations considered for each particular size.  Since the
number of Goldstone modes exactly equals the number of different
realizations of disorder considered, their peaks are in a ratio of
roughly 200/100/50 with respect to one another. The main observation
is that the histograms for the three sizes are very similar, with a
huge peak just below IPR=1. The absence of a dependence on $N_d$
confirms that all these modes are localized (except the Goldstone
modes, of course).

\begin{figure}
\centering
\includegraphics[angle=270,width=\FigWidth]{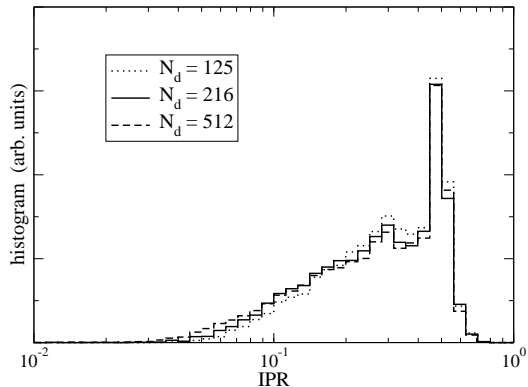}
\caption{\label{fig10} Histogram of the IPR values for all spin-waves
(spin flips) of energy $E > 10^{-1})$, for $N_d = 125, 216$ and 512.
Results are shown only for strong disorder configurations.  The
histograms are almost identical, suggesting only localized modes at
these energies.  }
\end{figure}

The high-energy localized states are of very different nature: their
 IPR is close to $0.5$, suggesting that it has large contributions
 from only two sites. Analysis of the $Y_i$ values for such modes
 shows that most of them are associated with nearest-neighbor (n.n.)
 Mn spins on the FCC Ga sublattice. The characteristic energy for the
 spin-waves centered on such n.n. sites is just below 5 meV, and this
 is the feature responsible for the peak appearing in the DOS at these
 energies (see Fig. \ref{fig7}, lower panel). Careful inspection of
 the DOS reveals the appearance of smaller peaks at somewhat lower
 energies (even the DOS for moderate disorder has a small peak at
 around 2 meV, and its IPR for these modes is close to 0.5). These
 peaks are associated with excitations of spin-pairs (or larger
 clusters) with varying separations, but they are not as well defined
 as the one corresponding to nearest-neighbors spins.

The small clusters of Mn spins giving rise to such modes are always
found to be in regions densely populated with charge carriers. Due to
their closeness in space, these spins are much more strongly coupled
to one another than they are to other spins with which they share
charge carriers. This leads to the ``resonance-like'' character of
these modes: either of the spins can be flipped with equal
probability.  As a result of the strong-coupling, the cost of flipping
either spin is high, since it frustrates their FM arrangement. Indeed,
the characteristic energy of roughly $J/3$ reflects much stronger
coupling than the average one present in the superlattice case. The
histogram of the IPR of all modes of energy $E > 10^{-1})$ shown in
Fig. \ref{fig10} confirms all these conclusions. We caution that these
energies may be substantially modified due to direct antiferromagnetic
Mn-Mn exchange (left out in our model) in real systems.

\begin{figure}
\centering
\includegraphics[width=\FigWidth]{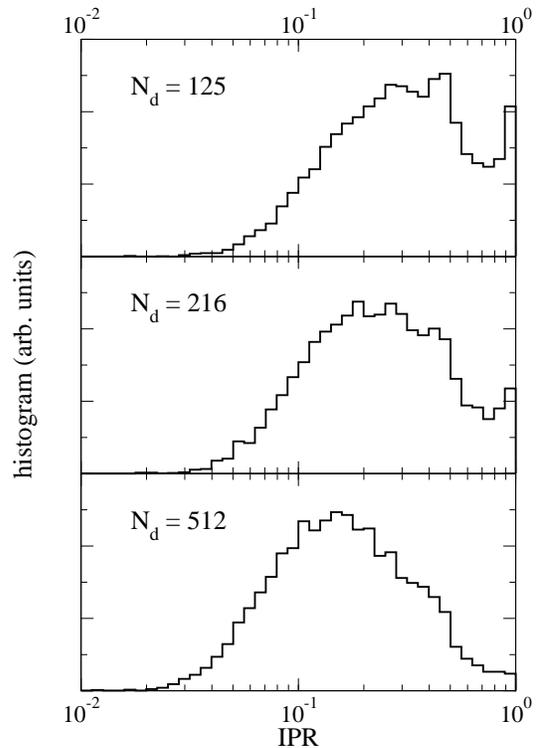}
\caption{\label{fig11} Histogram of the IPR values for all spin-waves
(spin flips) of energy $ 5\cdot 10^{-3})< E < 10^{-1})$, for $N_d =
125$ (upper), 216 (middle) and 512 (lower panel).  Results are shown
only for strong disorder configurations. With increasing $N_d$ the
histograms shift to lower values, suggesting that these spin-wave
modes are extended.  }
\end{figure}

Finally, the spin-waves at intermediate energies $5\cdot 10^{-3} < E <
10^{-1}$ correspond to extended modes. It is rather difficult to
establish exactly the ``mobility edge'' corresponding to the
transitions from localized to extended modes at either end, given the
rather small system sizes considered here. Thus, we use the values
quoted above only as plausible estimates of these boundaries.
Clearly, for these energies the average IPR decreases with increasing
$N_d$ (see Fig. \ref{fig8}).  In Fig. \ref{fig11} we plot the
histograms of the IPR for all the modes with energy $ 5\cdot 10^{-3}<
E < 10^{-1}$. In this case, size dependence is clearly visible, both
in the position of the broad peak maximum as well as in the lower
cut-off in IPR value. The position of the maxima scale roughly like
$1/N_d$. We have attempted both gaussian and lorentzian fits, but
neither seems to capture the low-IPR tail properly. In the smaller
samples, a second peak appears just below IPR=1. At first sight, this
might seem to indicate the existence of a finite density of localized
states at these energies as well. However, we believe that this peak
is a finite-size effect. Clearly, its amplitude decreases with
increasing $N_d$ and would vanish in the thermodynamic limit. This is
very different from the behavior observed in either of the two
localized energy ranges, where the distributions for all sizes are
virtually identical (see Figs. \ref{fig9} and \ref{fig10}).  The
reason for the appearance of this second peak is simply the
restriction IPR $\le$ 1. As the broad peak moves towards higher values
with decreasing $N_d$, all the values in the upper tail ``bunch'' at
IPR=1.

One final important observation relates to the absolute values of the
IPRs in the extended spin-wave regime. Although scaling with system
size is present, the corresponding IPR values are much higher than the
ones expected for modes extended over {\em all} Mn sites (these values
are shown by the positions of the Goldstone modes in Fig. \ref{fig9},
and are well below the cut-offs observed in Fig. \ref{fig11}). This
means that in the disordered system, the extended spin-waves are
delocalized only over a fraction of the total number of sites. As the
amount of disorder increases, this fraction decreases, since IPR
averages for the moderate disorder are smaller than IPR averages for
full disorder.

These results are consistent with the picture provided by the
temperature dependent mean-field study of this model
(Ref. \onlinecite{MB2}). There, we concluded that in the disordered
system, the (small fraction $p$) of charge carriers are concentrated
in a small volume of the sample, where the local Mn concentration is
larger than average. As a result, the concentration of charge carriers
is strongly enhanced in these regions, and the exchange with the Mn
spins inside these regions (which is proportional to the probability
of finding charge carriers nearby) is greatly increased, preserving
magnetization of these regions up to high temperatures.  On the other
hand, the Mn spins in the regions devoid of charge carriers are very
weakly coupled, and behave as free spins down to very low
temperatures. Due to the very inhomogeneous assembly of weakly- and
strongly- interacting spins, the magnetization curves have very
unusual, concave shapes.

The present study of the spin-waves corroborates the same picture. We
find very low-energy, spin-flip like excitations, which are obviously
due to the weakly interacting spins, and which are responsible for a
very sharp decrease of the magnetization at exponentially small
temperatures.\cite{MB1,MB2} This is is to be contrasted with the
behavior of ordered conventional ferromagnets, where at low
temperatures only low-energy long-wavelength spin-waves can be
excited. Since their phase space is vanishing in the long-wavelength
limit ($\sim k^d$ in $d$ dimensions), the magnetization of
conventional ferromagnets decreases very slowly from its saturation
value with increasing temperature, leading to convex upwards
magnetization curves.\cite{Kaneyoshi}

In our model, the extended spin-waves are concentrated around the
high-density Mn regions, where the charge carriers mediating the
interactions are to be found. This is consistent with the appearance
of modes whose IPR, while scaling with the system size, shows modes
extended only over a fraction of all system sites. As the amount of
disorder decreases all the way to a fully ordered superlattice, the
charge carriers are more and more homogeneously spread throughout the
sample and the IPR of the extended modes has lower and lower values,
as observed from Fig. \ref{fig8}. However, the more homogeneous a
sample is, the less the average AFM coupling between the Mn spins and
the charge carriers, since the charge carriers have now equal
probability of being found anywhere in the sample, instead of being
concentrated in a small fraction of the space. This leads to a
decrease of the critical temperature $T_C$, as observed in both
mean-field \cite{MB1} and Monte-Carlo\cite{malcolm2} analysis.  The
enhancement of the mean-field $T_C$ with increasing disorder is
suggested also by the existence of the high-energy localized modes,
which show that ferromagnetic alignment will persist in high-density
clusters up to very high temperatures. It also suggests that local
ferromagnetic fluctuations might be observed well above $T_C$.

\section{Final Remarks}
\label{sec6}

The aim of this paper is two-fold. First, we demonstrate the accuracy
and speed of a new scheme for computing RPA spectra, which allows
tackling of systems of much larger sizes than the ones that can be
analyzed with the standard RPA formulation. Investigation of large
system sizes and averages over many disorder realizations facilitate a
clear picture for the problem we are interested in, namely the
spectrum and nature of spin-waves of a disordered DMS.

We then demonstrate that disorder can significantly change the
spectrum and the nature of the spin-waves. This is likely to lead to
important consequences not only as far as magnetic properties are
concerned (we have already commented on the fast de-magnetization with
increasing temperature, due to low-energy spin-flip modes). More
importantly, this may have significant consequences for transport
properties as well. For instance, charge carrier spin scattering is
likely to be very different in various regions of the sample. Large
anomalous Hall effect has been observed in (Ga,Mn)As,\cite{Ohnorev}
but the theory used to interpret it is borrowed from phenomenology
relevant to homogeneous ferromagnetic metals. In an inhomogeneous
system, some of the accepted ideas might have to be modified or at
least verified to still hold true.

\section*{Acknowledgments}
This research was supported by NSF DMR-9809483.  M.B. was supported in
part by a Postdoctoral Fellowship from the Natural Sciences and
Engineering Research Council of Canada.

\end{document}